\documentclass[%
 reprint,
superscriptaddress,
 amsmath,amssymb,
 aps,
prb,
]{revtex4-2}
\usepackage{graphicx}
\usepackage{dcolumn}
\usepackage{bm}
\usepackage{siunitx}
\usepackage{kantlipsum}
\usepackage{xcolor}
\usepackage{appendix}
\usepackage{filecontents}

\begin{document}
\title{ Double anticrossings induced by nonlinear magnon interactions}
\author{Geil Emdi}
\affiliation{RIKEN Center for Emergent Matter Science (CEMS), Wako 351-0198, Japan}
\author{Tomosato Hioki}
\affiliation{RIKEN Center for Emergent Matter Science (CEMS), Wako 351-0198, Japan}
\affiliation{Department of Applied Physics,The University of Tokyo,Tokyo 113-8656, Japan}
\affiliation{Advanced Institute for Materials Research, Tohoku University, Sendai 980-8577, Japan}
\author{Koujiro Hoshi}
\affiliation{Department of Applied Physics,The University of Tokyo,Tokyo 113-8656, Japan}
\affiliation{Advanced Institute for Materials Research, Tohoku University, Sendai 980-8577, Japan}
\author{Takahiko Makiuchi}
\affiliation{RIKEN Center for Emergent Matter Science (CEMS), Wako 351-0198, Japan}
\author{Aoi Yamauchi}
\affiliation{Department of Applied Physics,The University of Tokyo,Tokyo 113-8656, Japan}
\author{Eiji Saitoh}
\affiliation{RIKEN Center for Emergent Matter Science (CEMS), Wako 351-0198, Japan}
\affiliation{Department of Applied Physics,The University of Tokyo,Tokyo 113-8656, Japan}
\affiliation{Advanced Institute for Materials Research, Tohoku University, Sendai 980-8577, Japan}
\affiliation{Institute for AI and Beyond, The University of Tokyo,Tokyo 113-8656, Japan}
\begin{abstract}
We observe pump-induced double anticrossings whose gap sizes and center-frequency shifts depend strongly on pump power, indicating a nonlinear mechanism. The anticrossings vanish at high magnetic fields, where energy conservation suppresses three-magnon splitting, thereby identifying the underlying process as three-magnon scattering. We attribute the double anticrossings to nondegenerate three-magnon splitting, which generates two magnon populations at distinct frequencies. Each population forms a standing-wave mode and couples independently to the Kittel mode, giving rise to two effective coupling channels. These results demonstrate that nonlinear magnon interactions can dynamically generate multiple coupling channels within a single system.  
\end{abstract}
\maketitle

\begin{figure}
    \centering
    \includegraphics[width=1\linewidth]{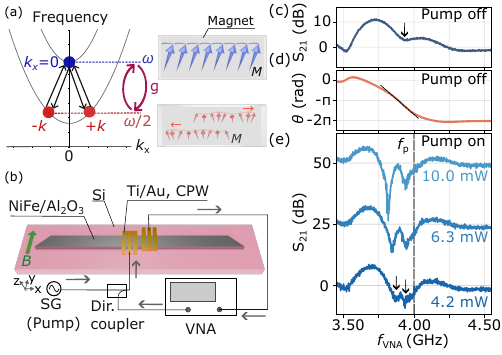}
    \caption{(a) Schematic of the spin wave dispersion branches and  three-magnon splitting and confluence processes giving rise to an effective coupling $g$. $M$ denotes the magnetization.
(b) Experimental setup (SG: signal generator; VNA: vector network analyzer). 
(c) Magnitude and  (d)  phase ($\theta$) of the transmission $S_{21}$ at $B = \SI{13.1}{\milli\tesla}$ with pump off condition. The dashed line indicate slope near the resonant mode frequency  (e) $S_{21}$ magnitude at pump on condition with $f_{\mathrm{p}} = \SI{4}{\giga\hertz}$ and various pump power $P$.}
    \label{fig:placeholder1}
\end{figure}

Coupling between oscillators underpins a wide range of physical systems, from atomic ensembles to condensed-matter platforms \cite{Araujo2016Superradiance,Boller1991EIT,Abdumalikov2010PRL,Thompson1992NormalMode,Reithmaier2004StrongCoupling,SafaviNaeini2019Optica,Lee2015,Tabuchi2015,Shen2022PRL,Hioki2022CommPhys,Gustafsson2014Science}. At its core, coupling between two modes hybridizes them into new eigenmodes separated by a frequency gap — observed as an anticrossing (avoided crossing) in a spectrum \cite{Devaquet1978,Sillanpaa2009,Saglamyurek2018,Zhang2014}. Extending this concept to three or more coupled modes leads to rich spectral phenomena, including bright–dark mode manifolds \cite{ Li2022RemoteMagnon, Gabor2025EPJQT, RubiesBigorda2022PRR, Zhang2015Dark, Zhan2021BrightDark} and higher-order exceptional points in non-Hermitian systems \cite{Wang2019NatCommun, Ozdemir2019NatMater, Heiss2012JPA}. In the linear regime, these phenomena arise from coupling channels that are fixed by design, where the number of channels is set by the number of resonators and their connections \cite{Li2022RemoteMagnon,Zhang2015Dark, Zhan2021BrightDark,Roy2018PRA}. 

Nonlinear dynamics offer an alternative route, enabling new modes and coupling channels to dynamically emerge within a system \cite{Menotti2019PRL,Clerk2020Hybrid,BoydNonlinearOptics,Aspelmeyer2014RMP}. In magnets, magnons—quanta of collective spin excitations that propagate as spin waves—exhibit strong intrinsic nonlinearities  \cite{Suhl1957,Rezende1986,Lee2023} that enable three-magnon splitting and confluence processes: a driven magnon decays into two lower-frequency magnons, while the reverse confluence process recombines them into the driven excitation \cite{Suhl1957,Irvine2003,Bauer2015,Okano2019,Shimizu2022,Qu2023,Makiuchi2024}, as shown in Fig. 1(a). These processes enable effective coupling between the driven and lower-frequency magnon populations \cite{Qu2025PumpInduced,Sud2025ElectricalNonlinear,Arfini2025NonlinearMagnon}. So far, most experiments have employed uniform excitation (wavevector, $k=0$)  to drive the three-magnon processes, where momentum conservation restricts splitting to counterpropagating $\pm k$ modes with equal frequency. Thus, the lower-frequency magnons occupy frequency-degenerate states, yielding only one effective coupling channel \cite{Sud2025ElectricalNonlinear,Arfini2025NonlinearMagnon}. 
Multiple coupling channels have been reported when several pre-existing magnetostatic standing-wave modes are simultaneously present \cite{Qu2025PumpInduced}; but in that case, each channel originates from an independent eigenmode defined by the device geometry rather than from the nonlinear dynamics themselves. This motivates the question of whether nonlinear dynamics alone can generate multiple coupling channels within a physical system, without relying on multiple pre-engineered resonant modes.

\begin{figure*}
    \centering
    \includegraphics[width=1\linewidth]{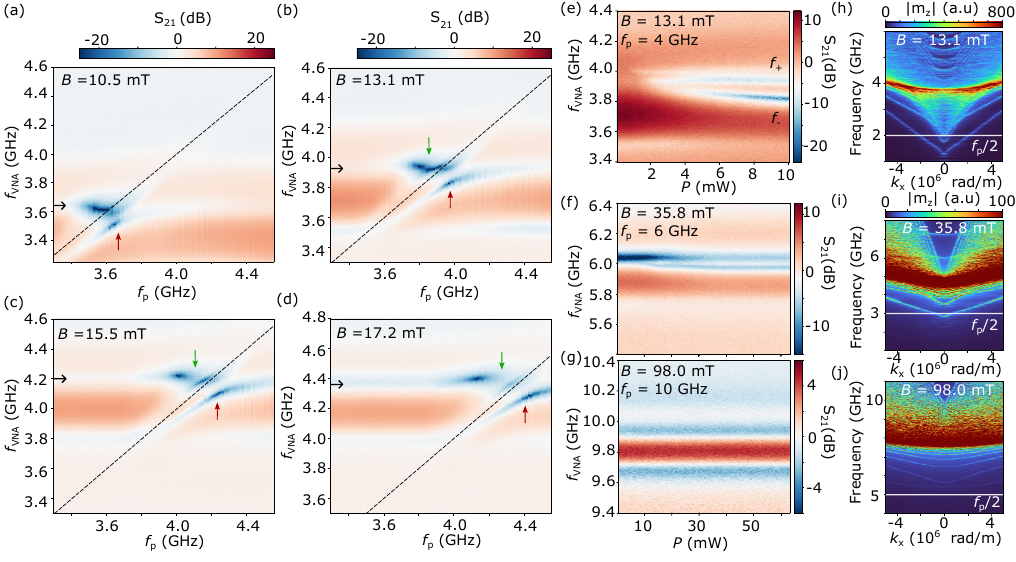}
    \caption{Pump-frequency dependence of the $S_{21}$ spectrum for different magnetic fields: (a) 10.5 mT, (b) 13.1 mT, (c) 15.5 mT, and (d) 17.2 mT at $P = \SI{8}{\milli\watt}$. Black dashed lines indicate $f_{\mathrm{VNA}} = f_{\mathrm{p}}$, and black arrows mark the Kittel mode frequencies. Power-dependent of $S_{21}$ at (e)  13.1 mT with  $f_\mathrm{p}= 4$ GHz, (f) at 35.8 mT with $f_\mathrm{p}= 6$ GHz, and (g) at 98 mT with $f_\mathrm{p}= 10$ GHz. Calculated spin-wave dispersions at (h) 13.1,  (i) 35.8, and (j) 98 mT, where $|m_z|$ denotes the z-component (out-of plane) of the magnetization in the sample simulation. The  colorbar used in panel (j) is the same as that of panel (i).}
    \label{fig:placeholder2}
\end{figure*}

Here, we demonstrate that nonlinear interactions can generate multiple emergent coupling channels within a single system. Using microwave pumping which predominantly excite finite-wavevector magnons, we observe pump-induced double anticrossings, whose frequency gaps and central frequency shifts exhibit a pronounced dependence on the pump power. The anticrossings disappear at high magnetic fields, where energy conservation forbids three-magnon splitting, strongly supporting their three-magnon origin. We attribute the observed double anticrossings to nondegenerate three-magnon processes mediated by finite-wavevector pumped magnons. Furthermore, the measured power-law scaling of both the frequency gaps and the central frequency shifts is in good agreement with our proposed mechanism.

We  sputtered a  \SI{50}{\nano\meter}-thick permalloy (NiFe)  film onto a silicon (Si) substrate and patterned it into a trapezoidal geometry with a width of \SI{50}{\micro\meter} and side lengths of \SI{740}{\micro\meter} and \SI{640}{\micro\meter} by photolithography and lift-off process. A \SI{100}{\nano\meter}-thick \(\mathrm{Al}_2\mathrm{O}_3\) layer is sputtered on top of Py for electrical insulation, followed by \SI{20}{\nano\meter}-thick titanium (Ti) adhesion layer and \SI{300}{\nano\meter}-thick gold (Au) films patterned to form two \SI{14}{\micro\meter}-width coplanar waveguides (CPWs) with \SI{80}{\micro\meter} separation. Microwaves are applied to one CPW using a signal generator supplying a continuous-wave pump at frequency \(f_{\mathrm{p}} = \frac{\omega_\mathrm{p}}{2\pi}\) and a vector network analyzer (VNA) providing a weak swept probe at VNA frequency ($f_{\mathrm{VNA}}$) to measure transmission  \(S_{21}\) spectrum across the two CPWs. We apply a static magnetic field \(B\) in-plane, along the y-axis indicated in Fig. 1(b). 

Before presenting the main data, we measure the transmission between the two CPWs without a microwave pump at $B=\SI{13}{\milli\tesla}$, normalized to a reference $S_{21}$ taken at $B=\SI{200}{\milli\tesla}$. The magnitude and phase of $S_{21}$ are shown in Fig.~1(c) and (d).
The spectrum exhibits a broad background with a superimposed dip at $f_{\mathrm{VNA}} = 3.95~\mathrm{GHz}$ (black arrow). The phase shows a rapid variation near this frequency consistent with a resonant response \cite{Weiss2022Exp}. We attribute this dip to a standing wave resonance, while the broad background arises from propagating spin waves. Under microwave pumping at $f_{\mathrm{p}}=\SI{4}{\giga\hertz}$, this single dip splits into two distinct dips [Fig.~1(e)]. As the pump power $P$ increases, the separation between the dips grows and their central frequency shifts, indicating nonlinear dynamics involving the standing wave resonance.

We present the $S_{21}$ spectra as a function of VNA frequency and pump frequency at selected magnetic fields [Fig.~2(a--d)]. At low field, a single anticrossing appears below the pump frequency $f_\mathrm{p}$, associated with the standing wave resonance [Fig.~2(a), red arrow]. With increasing field, an additional gap emerges above $f_\mathrm{p}$, initially narrow [Fig.~2(b), green arrow] and progressively widens [Fig.~2(c)], resulting in a clear double anticrossing at $B=\SI{17.2}{\milli\tesla}$ [Fig.~2(d)]. In all cases, the  anticrossings track the standing wave resonance [Fig. 2(a)-(d), black arrows], whose frequency shifts with magnetic field, confirming its magnonic origin. Micromagnetic simulations further identify this standing wave resonance as the Kittel mode ($k \approx 0$)~\cite{SI}. The evolution from a single to a double anticrossing indicates that an additional coupling channel becomes active with increasing field, suggesting the presence of two distinct coupling channels involving the Kittel mode.

To confirm that the anticrossings originate from three-magnon processes, we examine their power and field dependence. At $B = \SI{13.1}{\milli\tesla}$ and $f_\mathrm{p} = \SI{4}{\giga\hertz}$, a single dip splits into two modes at $f_\pm$ once the pump power exceeds a threshold $P_{\mathrm{th}}$ [Fig.~2(e)]. With increasing power $P$, the frequency gap $f_{\mathrm{gap}} = f_+ - f_-$ widens, while the central frequency $(f_+ + f_-)/2$ shifts away from the pump frequency $f_\mathrm{p}$. At $B = \SI{35.8}{\milli\tesla}$ [Fig.~2(f)], $f_{\mathrm{gap}}$ are narrower than at 13.1~mT. Upon further increasing the field to $B = \SI{98}{\milli\tesla}$, $f_{\mathrm{gap}}$ disappears entirely [Fig.~2(g)], with no observable frequency shift or spectral distortion. 

This behaviour is understood from energy conservation: three-magnon splitting requires spin-wave states to exist at or below $f_\mathrm{p}/2$. Increasing the magnetic field shifts the spin-wave band to higher frequencies, progressively reducing the available phase space for splitting until it is entirely suppressed. This is supported by the calculated spin-wave dispersion relations shown in Figs.~2(h)--(j) \cite{SI}. At $B = \SI{98}{\milli\tesla}$, no spin-wave bands remain at or below $f_\mathrm{p}/2$, prohibiting three-magnon splitting. The simultaneous disappearance of both the frequency gaps and the spectral shifts at high field therefore supports the three-magnon origin of the anticrossings and rules out alternative nonlinear mechanisms such as Kerr-type effects.

\begin{figure}
    \centering
    \includegraphics[width=1\linewidth]{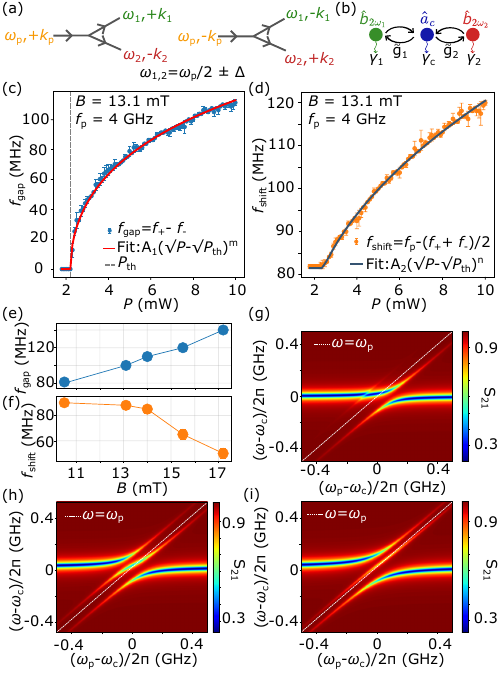}
    \caption{(a) Schematic illustrations for nondegenerate parametric excitation by the finite-wavevector pump magnon modes (b) Schematic of the effective coupling mechanisms included in the model.  (c) $f_\mathrm{gap}$ vs pump power $P$. Solid line: fit to 
$f_\mathrm{gap}=A_1(\sqrt{P}-\sqrt{P_\mathrm{th}})^{m}$ 
with $A_1=480$ MHz/W$^{m/2}$ and $m=0.49$.
(d) $f_\mathrm{shift}$ vs $P$. Solid line: fit to 
$f_\mathrm{shift}=A_2(\sqrt{P}-\sqrt{P_\mathrm{th}})^{n}$ 
with $A_2=500$ MHz/W$^{n/2}$ and $n=0.95$. Both taken at 13.1 mT with  $f_\mathrm{p}= 4$ GHz. (e) $f_\mathrm{gap}$ and (f) $f_\mathrm{shift}$   measured at magnetic fields of 10.5, 13.1, 14.0, 15.5, and 17.2~mT, using pump frequencies of 3.65, 4.00, 4.08, 4.20, and 4.38~GHz, respectively. (g--i) Simulated $S_{21}$ spectra from the three-mode coupled model for increasing coupling strengths and decreasing detuning: (g) $\tilde{g}_1 = 15~\mathrm{MHz}$, $\tilde{g}_2 = 50~\mathrm{MHz}$, $\Delta = 44~\mathrm{MHz}$; (h) $\tilde{g}_1 = 45~\mathrm{MHz}$, $\tilde{g}_2 = 60~\mathrm{MHz}$, $\Delta = 32.5~\mathrm{MHz}$; (i) $\tilde{g}_1 = 65~\mathrm{MHz}$, $\tilde{g}_2 = 65~\mathrm{MHz}$, $\Delta = 25~\mathrm{MHz}$. Parameters are related to experimental observables via $\tilde{g}_2 = f_{\mathrm{gap}}/2$ and $\Delta = f_{\mathrm{shift}}/2$. Fixed parameters used in all simulations are $\kappa = 50~\mathrm{MHz}$, $\kappa_\mathrm{ex} = 20~\mathrm{MHz}$, and $\gamma_{1,2} = 20~\mathrm{MHz}$. }
    \label{fig:placeholder3}
\end{figure}

We now develop a microscopic model to account for the double anticrossing and its key features. Owing to the localized field profile of the CPW, the pump microwave couples more efficiently to finite-wavevector magnons, predominantly exciting propagating spin waves with wavevector, $k_\mathrm{p} \neq 0$. These pumped modes undergo nondegenerate three-magnon splitting into secondary mode pairs with frequencies and wavevectors: $(\omega_1, k_1)$ and $(\omega_2, -k_2)$ where $k_1 \neq k_2$, satisfying $\omega_\mathrm{p} = \omega_1 + \omega_2$ and $k_\mathrm{p} = k_1 - k_2$, along with the corresponding counterpropagating process for the $-k_\mathrm{p}$ pumped mode [Fig. 3(a)], described by the non-degenerate three-magnon process Hamiltonian
\begin{equation}
H_\mathrm{nd}/\hbar
= V_p^{(3)} a_{k_\mathrm{p}}^\dagger a_{k_1} a_{-k_2}
+ V_p^{(3)} a_{-k_\mathrm{p}}^\dagger a_{-k_1} a_{k_2}
+ \mathrm{H.c.},
\end{equation}
where $a_{\pm k_\mathrm{p}}$ ($a_{\pm k_\mathrm{p}}^\dagger$) denote annihilation (creation) operators for the pumped modes, $a_{\pm k_1}$ and $a_{\pm k_2}$ correspond to the secondary magnons, and $V_p^{(3)}$ is the three-magnon coupling strength. Since $k_\mathrm{p} \neq 0$, momentum conservation enforces nondegenerate solutions ($k_1 \neq k_2$), generating two magnon populations at distinct frequencies $\omega_{1,2} = \omega_\mathrm{p}/2 \pm \Delta$, where $\Delta$ is the detuning parameter.

Within each population, counterpropagating modes $(k_j, -k_j)$ ($j = 1,2$) form standing-wave states that couple to the Kittel mode $(\omega_c, k_c \approx 0)$ via three-magnon confluence and splitting processes, while satisfying momentum conservation $k_c = k_j + (-k_j) \approx 0$. These interactions are enabled by their spatial overlap with the Kittel mode and are described by the coupling Hamiltonian
\begin{equation}
H_\mathrm{c}/\hbar
= V_1^{(3)} a_c^\dagger a_{k_1} a_{-k_1}
+ V_2^{(3)} a_c^\dagger a_{k_2} a_{-k_2}
+ \mathrm{H.c.},
\end{equation}
where $a_c$ ($a_c^\dagger$) denotes the Kittel mode and $V_{1,2}^{(3)}$ are the coupling strengths. Expressing the magnon pairs in the standing-wave basis  $b_{j+}=(a_{k_j}+ i a_{-k_j})/\sqrt{2}$ and $b_{j-}=(i a_{k_j}+ a_{-k_j})/\sqrt{2}$, and applying a semiclassical mean-field approximation~\cite{Arfini2025NonlinearMagnon, SI} yields the beam-splitter type Hamiltonian
\begin{equation}
H_\mathrm{c,eff}/\hbar
\approx \sqrt{2} V_1^{(3)} \beta_1 \, a_c^\dagger b_{1+}
+ \sqrt{2} V_2^{(3)} \beta_2 \, a_c^\dagger b_{2+}
+ \mathrm{H.c.},
\end{equation}
where $\beta_{1,2}$ are the steady-state amplitudes of the secondary magnons in the rotating frame at $\omega_\mathrm{p}/2$. The resulting $H_\mathrm{c,eff}$ describes the interaction between the two magnon populations and the Kittel mode, with effective couplings $g_{1,2} = \sqrt{2} V_{1,2}^{(3)} \beta_{1,2}$.  In the laboratory frame, the secondary magnon fields oscillate at $\omega_{1,2}$, giving time-dependent couplings $g_{1,2}(t) = \tilde{g}_{1,2} e^{-i \omega_{1,2} t}$. Transforming to the rotating frames at $\omega_{1,2}$ via $b_{j+}(t) = {b}_{2\omega_j} e^{i \omega_j t}$ removes the oscillating terms and yields a time-independent three-mode model, where $b_{2\omega_{1},2\omega_{2}}$ represents the rotating-frame operator of the secondary magnons \cite{SI}. In this picture, the Kittel mode $a_c$ couples simultaneously to two magnon-pair modes ${b}_{2\omega_1}$ and ${b}_{2\omega_2}$ with static coupling strengths $\tilde{g}_{1,2}$ [Fig.~3(b)]. Since $\omega_1 \neq \omega_2$, the resonance conditions $2\omega_1 \simeq \omega_c$ and $2\omega_2 \simeq \omega_c$ are satisfied separately, producing two distinct coupling channels and hence the double anticrossing observed in $S_{21}$.

To experimentally test this theoretical model, we determine the steady-state magnon amplitudes $\beta_{1,2}$ by considering the total Hamiltonian $\mathcal{H}=H_0+H_{\mathrm{p}}+H_{\mathrm{s}}+H_{\mathrm{nd}}+H_{\mathrm{c}}$. The free Hamiltonian is given by $H_0/\hbar=\omega_c a_c^\dagger a_c+\omega_p\sum_{\sigma=\pm} a_{\sigma k_\mathrm{p}}^\dagger a_{\sigma k_\mathrm{p}}+\sum_{j=1,2}\omega_j\sum_{\sigma=\pm} a_{\sigma k_j}^\dagger a_{\sigma k_j}$. The coherent microwave pump driving the magnon modes at $\pm k_\mathrm{p}$ is described by $H_{\mathrm{p}}/\hbar=i\sum_{\sigma=\pm}(\varepsilon_p e^{-i\omega_p t}a_{\sigma k_\mathrm{p}}^\dagger-\mathrm{H.c.})$, where $\varepsilon_p$ is the pump amplitude. The weak probe field coupled to the Kittel mode is given by $H_{\mathrm{s}}/\hbar=\sqrt{\kappa_{\mathrm{ex}}}(s_{\mathrm{in}}a_c^\dagger+s_{\mathrm{in}}^\dagger a_c)$, where $s_{\mathrm{in}}$ is the input probe field and $\kappa_{\mathrm{ex}}$ is the coupling rate. From this model, we obtain the steady-state amplitudes $\beta_{1,2}$ as \cite{SI}
\begin{equation}
|\beta_{1,2}| =
\sqrt{
\frac{\sqrt{|\tilde{\gamma}_{2,1}|}}{|V_p^{(3)}|\sqrt{|\tilde{\gamma}_{1,2}|}}
\frac{|\tilde{\gamma}_{2,1}|}{\Re(\tilde{\gamma}_{2,1})}
\left(
|\varepsilon_p|
-
|\varepsilon_{\mathrm{th},0}|
\right)
},
\end{equation}
where $\tilde{\gamma}_{1,2}=\frac{\gamma_{1,2}}{2}+i\Delta_{1,2}-V_{1,2}^{(3)}|\tilde{a}_c|$. Here $\gamma_p$ and $\gamma_{1,2}$ denote the damping rates of the pumped and secondary magnon modes, respectively, and $\Delta_{1}=\Delta$, $\Delta_2=-\Delta$ are the frequency detunings, $\Re(\tilde{\gamma}_{1,2})$ is the real part of $\tilde{\gamma}_{1,2}$, and $\varepsilon_{\mathrm{th},0}=\gamma_p\sqrt{|\tilde{\gamma}_1||\tilde{\gamma}_2|}/(2|V_p^{(3)}|)$ is the parametric threshold.

These magnon amplitudes determine two experimentally accessible quantities. First, they set the effective couplings $g_{1,2}\propto\beta_{1,2}$ that govern the frequency gap. Second, they give rise to a pump-induced frequency shift through nonlinear self-energy corrections.
The frequency shift can be obtained from the steady-state Kittel mode amplitude $\tilde a_{\mathrm c}=-(s_{\mathrm{in}}+\mathcal N)/\mathcal D$, where \cite{SI}
\begin{equation}
\mathcal D
=
\left(i\Delta_c+\frac{\gamma_c}{2}\right)
-\frac{4|V_1|^2|\beta_1|^2}{i\Delta_1+\gamma_1/2}
-\frac{4|V_2|^2|\beta_2|^2}{i\Delta_2+\gamma_2/2},
\end{equation}
defines the effective susceptibility and $\mathcal N$ is an additional nonlinear source term \cite{SI}. The imaginary part of $\mathcal D$ yields a pump-induced frequency shift
\begin{equation}
\delta\omega_c
=
\frac{4|V_1|^2|\beta_1|^2\Delta_1}{\Delta_1^2+(\gamma_1/2)^2}
+
\frac{4|V_2|^2|\beta_2|^2\Delta_2}{\Delta_2^2+(\gamma_2/2)^2}.
\end{equation}
Thus, the model predicts (i) a spectral gap scaling as $f_{\mathrm{gap}}\propto g_{1,2} \propto \beta_{1,2}$, and (ii) a frequency shift scaling as $\delta\omega_c\propto|\beta_{1,2}|^2$. Importantly, the latter vanishes in the  limit $\Delta_{1,2}=0$, providing a direct signature of nondegenerate three-magnon splitting into magnons with distinct frequencies.

We experimentally test these predictions by measuring $f_{\mathrm{gap}}$ at $B=\SI{13.1}{\milli\tesla}$ and $f_\mathrm{p}=\SI{4}{\giga\hertz}$ as a function of pump power $P$ [Fig.~3(c)]. When one coupling dominates ($g_1 \ll g_2$), the gap is well approximated by
$f_{\mathrm{gap}}\approx2\sqrt{|g_2|^2-(\gamma_c-\gamma_2)^2/16}$, where $\gamma_c$ is the damping rate of Kittel mode \cite{SI}. As shown in Fig.~3(c), the measured gap follows $f_{\mathrm{gap}}\propto(\sqrt{P}-\sqrt{P_{\mathrm{th}}})^{0.49}$, consistent with the predicted scaling $g_{1,2}\propto\beta_{1,2}\propto\sqrt{\sqrt{P}-\sqrt{P_{\mathrm{th}}}}$, given that $\varepsilon_{p}\propto\sqrt{P}$. The extracted threshold $P_{\mathrm{th}}=\SI{2.3}{\milli\watt}$ reflects both $\varepsilon_{\mathrm{th},0}$  and damping contributions.

We test the second prediction by analyzing the shift of the spectral center $(f_+ + f_-)/2$ with increasing $P$ at $B=\SI{13.1}{\milli\tesla}$ and $f_p=\SI{4}{\giga\hertz}$ [Fig.~2(e)]. Defining $f_{\mathrm{shift}} = f_p - (f_+ + f_-)/2$, we obtain $|f_{\mathrm{shift}}| \propto (\sqrt{P}-\sqrt{P_{\mathrm{th}}})^{0.95}$ [Fig.~3(d)], in agreement with the predicted scaling $\delta\omega_c \propto |\beta_{1,2}|^2 \propto \sqrt{P}-\sqrt{P_{\mathrm{th}}}$, thus confirming the nondegenerate three-magnon splitting by the pump mode.

To understand the field evolution of the double anticrossings shown in Figs.~2(a)–(d), we extract $f_{\mathrm{gap}}$ and $f_{\mathrm{shift}}$ below the pump frequency at the point of minimal spectral gap [Figs.~2(a)–(d), red arrows]. As shown in Figs.~3(e) and (f), $f_{\mathrm{gap}}$ increases with magnetic field while $f_{\mathrm{shift}}$ decreases. The increase in $f_{\mathrm{gap}}$ reflects enhanced coupling as the parametrically excited magnons shift toward smaller wavevectors, where weaker two-magnon scattering reduces damping \cite{Kasuya1961Relaxation,Arias1999}. The decrease in $f_{\mathrm{shift}}$ tracks the reduction in frequency detuning, since $f_{\mathrm{shift}} \approx 2\Delta$, where $\Delta = (\omega_1 - \omega_2)/2$. To see why $\Delta$ decreases with field, we note that near $\pm k_{1,2}$ the parabolic dispersion gives $\omega_1 - \omega_2 \propto (k_1 - k_2)(k_1 + k_2)$. Since $k_1 - k_2 = k_\mathrm{p}$ is fixed by the pump, increasing the field reduces $k_1 + k_2$ and hence $\Delta$. This also naturally explains the observed asymmetry between the two anticrossing gaps: since $\Delta \neq 0$, the two populations lie at different wavevectors on the dispersion and experience different wavevector-dependent damping, yielding unequal coupling strengths $\tilde{g}_1 \neq \tilde{g}_2$ and hence unequal gaps. As the field increases and $\Delta \rightarrow 0$, both populations converge toward $\omega_\mathrm{p}/2$ on the dispersion, equalizing their damping and coupling strengths, consistent with the progressively more symmetric double anticrossing observed at high magnetic field in Fig.~2(d).

Using the experimentally extracted values of $f_{\mathrm{gap}}$ and $f_{\mathrm{shift}}$ to determine the coupling strengths $\tilde{g}_{1,2}$ and detuning $\Delta$, we simulate $S_{21}(\omega)$ using the three-mode coupled model \cite{SI}. In the simulation, the pump frequency $\omega_\mathrm{p}$ is varied while the secondary magnon-pair modes satisfy $2\omega_1 = \omega_\mathrm{p} + 2\Delta$ and $2\omega_2 = \omega_\mathrm{p} - 2\Delta$. The results are shown in Figs.~3(g)–(i) for increasing $\tilde{g}_{1,2}$ and decreasing $\Delta$, reproducing the key features of Figs.~2 (a) -(d). Specifically, larger $\tilde{g}_{1,2}$ widens the frequency gaps of both anticrossings, while decreasing $\Delta$ leads to the emergence of a central branch that becomes nearly parallel to $\omega = \omega_\mathrm{p}$ at small detuning, consistent with the experimental observation at $B=17.2$ mT [Fig.~2(d)].

In summary, we have demonstrated that finite-wavevector microwave pumping generates multiple coupling channels dynamically within a single magnetic system. Through nondegenerate three-magnon splitting, two magnon populations emerge at distinct frequencies, each forming a standing-wave state that couples independently to the Kittel mode, giving rise to pump-induced double anticrossings. The nonlinear origin of these anticrossings is quantitatively confirmed by the power scaling of both the frequency gap and the central frequency shift, and further supported by their suppression at high magnetic fields where energy conservation forbids three-magnon splitting. These results establish finite-wavevector nonlinear magnon dynamics as a route to dynamically generate multiple coupling channels within a single system, providing a new platform for exploring emergent phenomena such as bright–dark mode manifolds and higher-order exceptional points driven by nonlinear interactions rather than fixed by device geometry.

\begin{acknowledgments}
This work is partially supported by JST CREST (JPMJCR25A2, JPMJCR20C1, and JPMJCR20T2), JST PRESTO (JPMJPR24F9), JSPS KAKENHI (Grants No. JP22K14584 and JP25K17943), Advanced Technology Institute Research Grants.
\end{acknowledgments}

\end{document}